\documentclass[aps,preprint,superscriptaddress,showpacs]{revtex4}
\usepackage{graphicx}
\usepackage{subfigure}
\begin{document}
\preprint{Bozza}
\title{Recombination kinetics of a dense electron-hole
plasma in strontium titanate}
\author{A. Rubano}
\affiliation{Dipartimento di Scienze Fisiche, Universit\`{a} di
Napoli ``Federico II'', Complesso di Monte S.Angelo, v.\ Cintia,
80126 Napoli, Italy}
\author{D. Paparo}
\author{F. Miletto}
\affiliation{ CNR-INFM Coherentia, Complesso di
Monte S.Angelo, v.\ Cintia, 80126 Napoli, Italy}
\author{U. Scotti di Uccio}
\affiliation{DiMSAT, Universit\`{a} di Cassino, via Di Biase 43, 03043
Cassino (FR), Italy}
\affiliation{ CNR-INFM Coherentia, Complesso di
Monte S.Angelo, v.\ Cintia, 80126 Napoli, Italy}
\author{L. Marrucci}
\email{lorenzo.marrucci@na.infn.it} \affiliation{Dipartimento di
Scienze Fisiche, Universit\`{a} di Napoli ``Federico II'', Complesso
di Monte S.Angelo, v.\ Cintia, 80126 Napoli, Italy}
\affiliation{ CNR-INFM Coherentia, Complesso di
Monte S.Angelo, v.\ Cintia, 80126 Napoli, Italy}
\date{\today}
\begin{abstract}
We investigated the nanosecond-scale time decay of the blue-green light
emitted by nominally pure SrTiO$_3$ following the absorption of an intense
picosecond laser pulse generating a high density of electron-hole pairs.
Two independent components are identified in the fluorescence signal that
show a different dynamics with varying excitation intensity, and which can
be respectively modeled as a bimolecular and unimolecolar process. An
interpretation of the observed recombination kinetics in terms of
interacting electron and hole polarons is proposed.
\end{abstract}
\pacs{78.55.-m,78.47.+p,73.50.Gr,71.35.Ee}


\maketitle

\section{Introduction}
Transition-metal oxides with a perovskite-type structure form a
fascinating class of materials, with an extraordinarily varied
physics. Among them, SrTiO$_3$ (STO) is a prototype material, with a
simple cubic perovskite structure at room temperature. Its
electrical properties appear deceptively simple: it is a band-gap
insulator, with a very large dielectric constant due to the high
polarizability of its ionic lattice. Mobile charges can be however
added to STO by chemical doping or photon excitation, turning it
into a polaronic conductor\cite{yasunaga68,keroack84} and, at very
low temperatures, even into a superconductor.\cite{schooley64} STO
interfaces with other polar oxides have recently shown surprising
transport properties,\cite{ohtomo02,yamada04} in connection with
charge transfer through the interface and with polaron localization
effects.\cite{pentcheva06} Even in bulk STO, the precise nature of
the electron and hole polarons is still fairly
unclear.\cite{yasunaga68,keroack84,eagles96,stashans01,eglitis03,nasu03,grigorjeva04,yu05}
As for its optical properties, STO has recently gained renewed
attention following the discovery of a significant blue-light
photoemission from Ar$^+$-irradiated or $n$-doped STO
samples.\cite{kan05,kan06} This finding, besides its interest for
potential applications, adds to the rather puzzling luminescence
phenomenology of this
material.\cite{grabner69,aguilar82,leonelli86,hasegawa00,mochizuki05}
The photoluminescence from a material is also a direct probe of the
dynamics and recombination processes of its photoexcited charge
carriers.

Motivated by these considerations, in this work we investigated the
nanosecond-scale time-resolved fluorescence of nominally pure STO,
following excitation from an intense picosecond laser pulse
generating a ``plasma'' of electrons and holes having initial
densities well above $10^{20}$ cm$^{-3}$. To our knowledge, no study
probing this regime was reported hitherto, for STO or any other
perovskite-type oxide. To avoid any possible misunderstandings, we
should however emphasize that, in this paper, we are using the term
``dense plasma'' in a generic sense, i.e., to indicate a system with
a very large number of interacting electrons and holes that are, at
least initially, freely mobile (although it is possible, in our
case, that at least a fraction of them will localize a short time
after being generated), regardless of the strength of their
electrostatic interaction and of the clear manifestation of
many-body effects. It must be also noted that photoluminescence
spectroscopy in this regime, with a photogenerated carrier density
much higher than impurity concentration, is likely probing mainly
intrinsic properties of the material.

This article is organized as follows. Section II describes the
samples and the experimental procedure. The experimental results and
their modeling are reported in Section III. These results are then
discussed and tentatively interpreted within a microscopic framework
in Section IV. Section V contains some conclusive statements.

\section{Experiment}
We investigated five transparent stoichiometric STO crystals
produced by four different companies (all with specified impurity
levels below 150 ppm), in the form of 1 mm thick plates. All
measurements discussed in this paper were taken at room temperature.
In our photoluminescence experiments, the excitation was induced by
ultraviolet (UV) pulses having a wavelength of 355 nm, obtained as
third-harmonic of the output of a mode-locked Nd:YAG laser. The
corresponding excitation photon energy is 3.49 eV, well above the
indirect band-gap of STO (3.26 eV) and just above its
direct-transition edge (3.46 eV).\cite{cardona65,capizzi70} The UV
pulses had a duration of 25 ps (full-width at half-maximum), a
repetition rate of 10 Hz, and an energy of $0-2$ mJ. The UV beam was
weakly focused on the sample, with a spot size of $1.2{\pm}0.1$ mm
(radius at $1/e^2$ of the maximum). The resulting fluence in the
spot-center was therefore $0-90$ mJ/cm$^2$. Since at a wavelength of
355 nm the optical penetration depth is of about 1 $\mu$m and the
reflectance is about 25\%,\cite{cardona65,capizzi70} the estimated
peak (spot-center) energy density absorbed in the material surface
layer is $\approx0-600$ J/cm$^3$. This corresponds to a peak density
of photo-generated electron-hole ($e-h$) pairs of up to
$1.2{\times}10^{21}$ cm$^{-3}$. During all our experiments, no
visible photoinduced damage of the sample surfaces was induced and
no irreversible variation of the signal with time was seen.

\begin{figure}
\includegraphics[angle=0, width=0.48\textwidth]{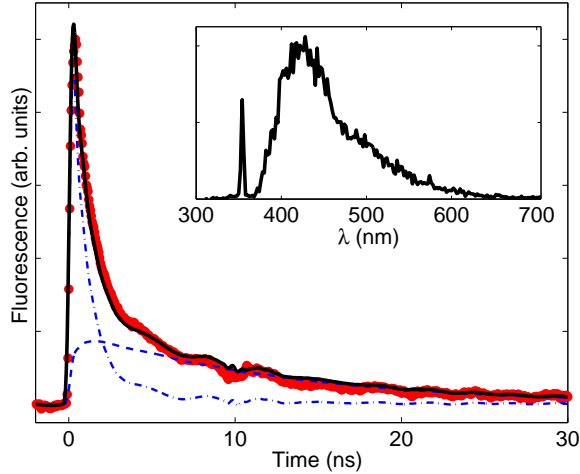}
\caption{(Color online) Example of photoluminescence signal. Main
panel: luminescence decay (pulse energy of 550 $\mu$J). Data are
shown as (red) filled circles, the black solid line is the
theoretical fit, taking into account the instrumental response (the
latter causes some minor signal features, such as the small wiggling
seen at 10 ns delay). The unimolecular (UD) and bimolecular (BD)
contributions (see text for a definition) are shown as (blue) dashed
and dot-dashed lines, respectively. Inset: emission spectrum (the
narrow peak seen at 355 nm is due to residual UV elastic
scattering).\label{fig1}}
\end{figure}

\begin{figure}
\includegraphics[angle=0, width=0.40\textwidth]{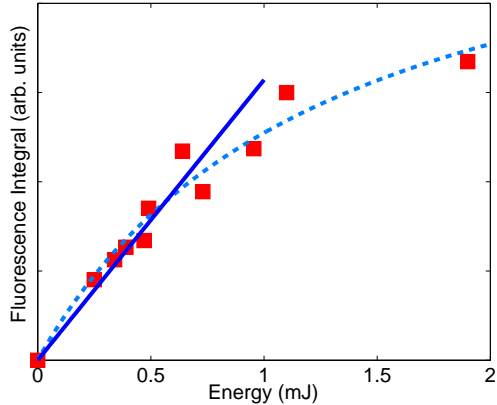}
\caption{(Color online) Photoluminescence integrated emission
intensity $I_{\mathrm{tot}}$ versus excitation pulse energy $E$.
Square dots are data. The solid line is a linear best-fit in the
energy range 0-1 mJ. The dashed line is a best-fit based on the
typical saturation behavior $I_{\mathrm{tot}}{\propto}E/(1+E/E_s)$,
that yields $E_s=1.3$ mJ.\label{fig2}}
\end{figure}

The luminescence emitted from the sample was collected by a lens
system imaging the illuminated sample spot onto the detector head,
after blocking the (much stronger) elastic scattering by a long-pass
filter with a cutoff wavelength of 375 nm. A typical
(time-integrated) luminescence spectrum is shown in the inset of
Fig.~\ref{fig1}. It peaks at 425 nm, in the blue, corresponding to
2.9 eV of photon energy, with a long tail reaching about 650 nm (1.9
eV). All samples, independent of the manufacturer, had a roughly
comparable fluorescence yield, with a maximum sample-to-sample
variation of 80\%. The initial emission intensity was found to be
even more stable, with a sample-to-sample variation below 30\%. This
stability is a clear indication that this luminescence is intrinsic,
although a dependence on intrinsic lattice defects (i.e., not
associated with impurities) cannot be excluded. Although we have not
performed absolute yield measurements, the photoluminescence is
clearly weak, indicating the dominance of non-radiative energy
relaxation. Finally, in all samples the spectrum profile and the
overall yield were found to be independent of excitation intensity,
with no significant sign of saturation, up to about 1 mJ of pump
pulse energies. We see instead a significant decrease of the yield
above 1 mJ, as shown in Fig.~\ref{fig2}. This extremely high
saturation intensity clearly cannot be ascribed to the filling of
impurity levels and therefore is a further confirmation of the
intrinsic nature of this photoluminescence. This saturation might be
instead related with the onset of the Auger recombination effect (we
will discuss this possibility later) or with plasma electromagnetic
screening effects (while we may probably exclude the contribution of
absorption ``bleaching'', which would increase the optical
penetration length and therefore reduce the absorbed energy density,
but would not reduce the overall absorbed energy and hence the
luminescence emission). The estimated plasma frequency of the
photogenerated carriers, assuming free electron masses for both
electrons and holes, is of about $6\times10^{14}$ Hz at the highest
excitation intensities, corresponding to an electromagnetic
wavelength of 500 nm. Therefore it is possible that some partial
screening of the UV excitation light is taking place at these
energies, thus decreasing the sample absorption efficiency. In the
rest of this paper, our analysis will be mainly focused on the
non-saturated energy range 0-1 mJ.

In time-resolved measurements, the luminescence was detected with a
photodiode (PD) having a rise-time of about 150 ps. In most
experiments the entire luminescence spectrum was integrated (with a
weight given by the PD detection efficiency), but in some cases we
inserted another filter in the detection line in order to select a
portion of the spectrum. The PD signal was acquired by a 20
Gsample/s digital oscilloscope having an analog bandwidth of 5 GHz.
With careful data analysis, taking into account the instrumental
response function (see below), this set-up has a time resolution of
about 100 ps. An example of a typical measured decay of the
luminescence is shown in Fig.~\ref{fig1}. The acquired signal $S(t)$
is actually the result of a convolution between the real
luminescence decay $I(t)$ and the detection-system response function
$r(t)$, where $t$ denotes time. The response function was measured
by detecting the elastic-scattered light of the excitation pulse.
Once $r(t)$ and $S(t)$ are known, the actual $I(t)$ decay may in
principle be calculated by a deconvolution procedure. However, in
order to avoid the typical problems associated with numerical
deconvolutions of data, we prefer here to compare the unprocessed
raw signals $S(t)$ with the theoretical luminescence decays
convoluted with $r(t)$.

\begin{figure*}
\centering \subfigure{
\includegraphics[angle=0, width=0.46\textwidth]{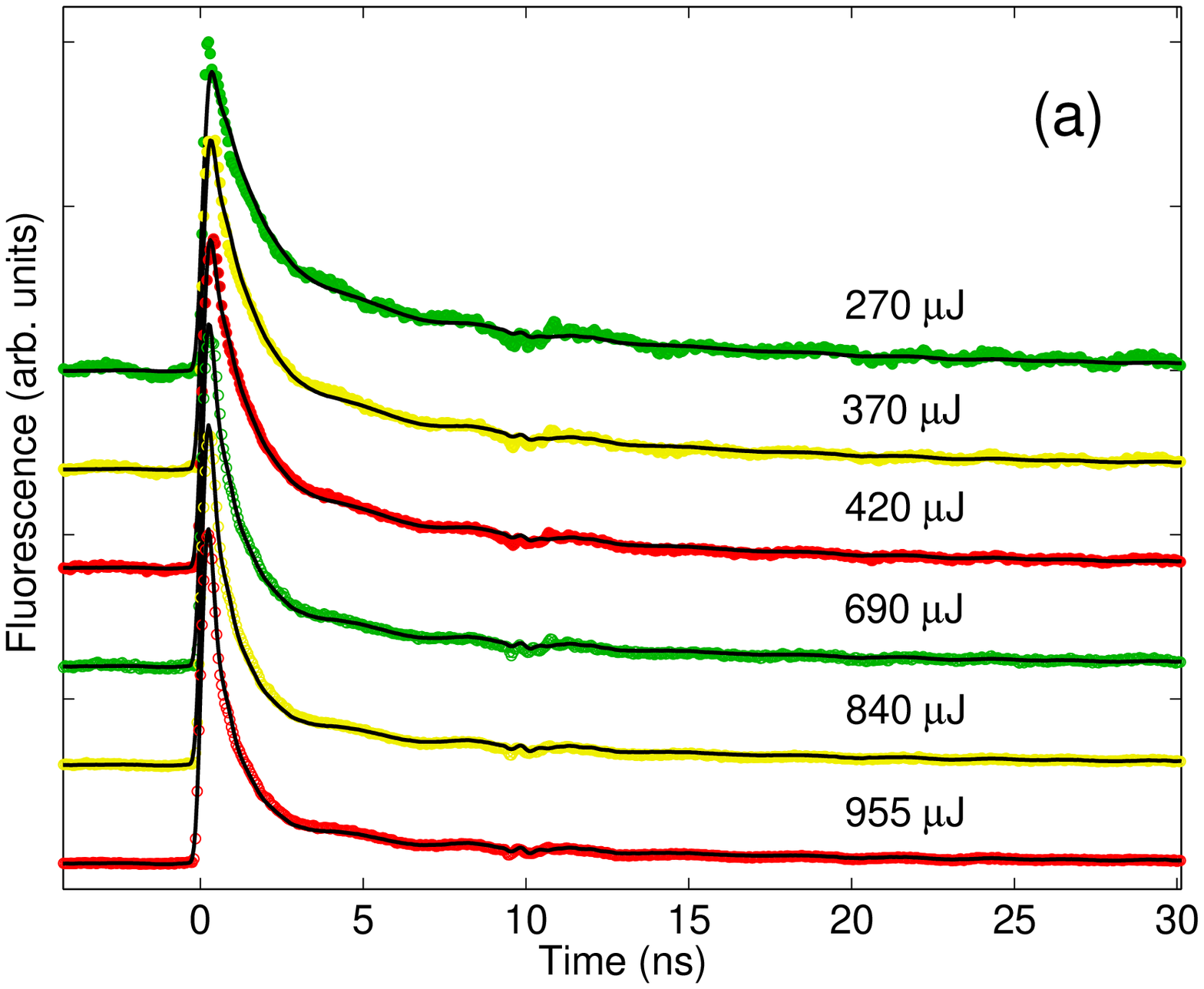}}
\subfigure{\includegraphics[angle=0,
width=0.485\textwidth]{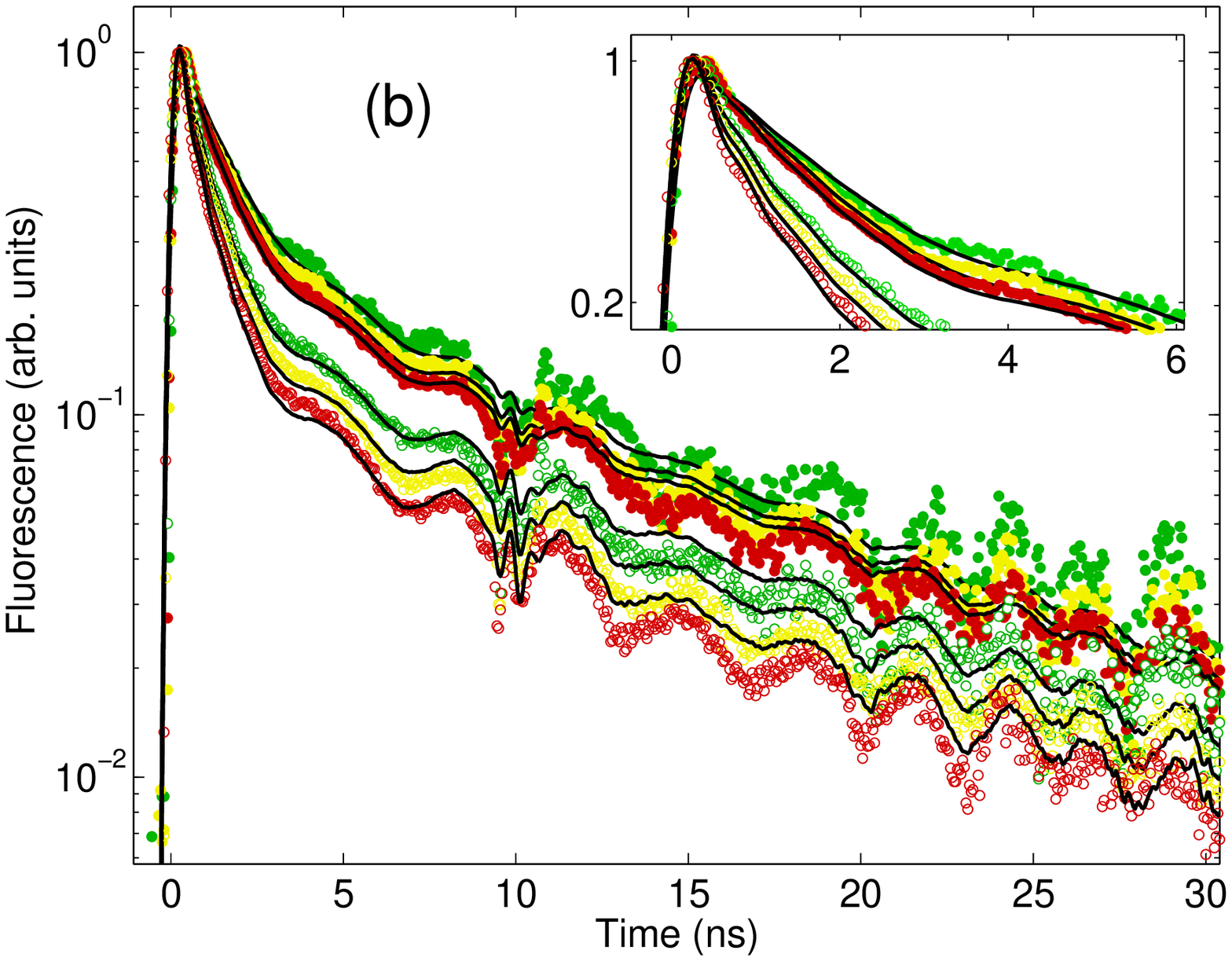}}

\caption{(Color online) Photoluminescence decay dependence on
excitation pulse energy (in the non-saturated range 0-1 mJ). Both
panels (a) and (b) show the same decay data (circles) referring to
one of the investigated samples, with the corresponding best-fit
curves (black solid lines) as emerging from a single global fit (see
text). Data and fits are normalized to their maximum. Panel (a) is
in linear scale and different decays are vertically offset for
clarity. Panel (b) is in semi-log scale, with no offsetting. The
inset in panel (b) is a zoomed-in view on the initial decay. The
oscillations in the best-fit curves seen in panel (b) result from
the convolution with the instrumental response function.
\label{fig3}}
\end{figure*}

\section{Results and analysis}
When varying the excitation intensity, the measured decay profile
changes markedly, as shown in Fig.~\ref{fig3}. In particular, the
initial decay becomes clearly faster for higher excitation
intensities (see also the inset in Fig.~\ref{fig3}b), while the
slower ``tail'' changes only its amplitude, but not its decay rate.
This dependence on excitation intensity excludes an interpretation
as a simple ``unimolecular'' exponential decay (UD) process and
implies that particle correlations are relevant. A ``bimolecular''
non-exponential decay (BD) naturally arises from a model in which
holes and electrons (or the corresponding polarons) have a
``partner-capture'' probability per unit time that is proportional
to the density of partners, and this capture event is also the
rate-limiting step for $e-h$ recombination (i.e., it may correspond
either to a direct recombination or to a recombination via an
intermediate short-lived state). Labeling with $N(t)$ the (equal)
density of photo-generated electrons and holes, the simplest ``pure
BD'' model is described by the following rate equation
\begin{equation}
\frac{dN}{dt}=-{\gamma}N^2(t), \label{bimol}
\end{equation}
where $\gamma$ is a constant. For excitation pulse energies below
the saturation level, the initial condition to be associated to this
equation is $N(0)={\alpha}U$, where $U$ is the excitation energy
density and $\alpha$ is a constant. We tried this model and found
that it provides an excellent fit to the energy dependence of the
initial faster decay but cannot account for the slower tail. An
obvious extension of the model is then to introduce also a UD
contribution in the decay. A unimolecular recombination arises when
there is decay channel in which the rate-limiting step is the
annihilation of $e-h$ bound pairs, while the binding dynamics is
much faster. There are different ways of combining UD and BD
processes together, and we tried some of the simplest possibilities
(see below). The model that gave the best results is based on the
assumption that there is a very rapid initial branching of the
charge carrier populations into two separate non-communicating
channels, respectively decaying with a UD and BD law. In other
words, we assume that, immediately after the UV excitation, two
separate charge-carrier populations of electrons and holes are
formed, with densities $N_1(t)$ and $N_2(t)$ (hole and electron
densities remain equal in each population). We further assume that
this initial separation has a fixed branching ratio, so that, for
any excitation energy density $U$, at time $t=0$ we have
$N_1(0)=\alpha_1U$ and $N_2(0)=\alpha_2U$, with $\alpha_1$ and
$\alpha_2$ constants (this is valid only for excitation energies
below the saturation level). Next, we assume that $N_1$ is governed
by a standard UD rate equation, $dN_1/dt=-N_1/\tau_1$, where
$\tau_1$ is an intensity-independent characteristic decay time. For
$N_2$ we assume that Eq.~(\ref{bimol}) remains valid. Finally we
introduce two different fluorescence quantum yields $Q_1$ and $Q_2$
for the two channels. This model leads to the following predicted
luminescence decay:
\begin{equation}
I(t)=-Q_1\frac{dN_1}{dt}-Q_2\frac{dN_2}{dt}=c_1 e^{-t/\tau_1} +
\frac{c_2}{(1+t/\tau_2)^2}, \label{model}
\end{equation}
where we have introduced the UD and BD amplitude coefficients
\begin{eqnarray}
c_1&=&Q_1N_1(0)/\tau_1=Q_1\alpha_1U/\tau_1\nonumber\\
c_2&=&Q_2{\gamma}N_2^2(0)=Q_2\gamma(\alpha_2 U)^2, \label{coeff}
\end{eqnarray}
and the (excitation-energy dependent) characteristic BD time
$\tau_2=1/({\gamma\alpha_2}U)$. Expression (\ref{model}) (after
convolution with the response function) provides an excellent fit to
the data, as shown for example in Fig.~\ref{fig1}, where the UD and
BD terms are also drawn separately. Further validation of our model
was obtained by requiring that Eq.~(\ref{model}) should successfully
fit the whole data set collected as a function of excitation energy,
with a single choice of the adjustable parameters $(\gamma\alpha_2)$
and $\tau_1$. We note that such parameters determine the BD and UD
decay times for each given excitation energy $U$, and are therefore
independently constrained by each single decay. The excellent fits
reported in Fig.~\ref{fig3} demonstrate that $(\gamma\alpha_2)$ and
$\tau_1$, corresponding in our model to material properties, are
indeed independent of excitation energy, thus providing a strong
confirmation of our model validity. This same global fit procedure
allowed to rule out alternative models.

Besides the successful model, we tried the following alternative
ones: (i) a pure BD equation such as Eq.~(\ref{bimol}), but acting
on unbalanced electron-hole populations (e.g., $N_e(t)=N_h(t)+C$),
as could arise due to trapping or impurity doping; (ii) a balanced
undivided population $N(t)$ of electrons and holes obeying the
single rate equation $dN/dt=-{\gamma}N^2-N/\tau_1$. The first model
cannot fit the single decay. The second can fit the single decay,
but not the whole set.

Moreover, we tested a model in which a trimolecular-decay process of
the kind $dN/dt=-C_AN^3$, such as the Auger effect, takes place in
the initial stage of the decay, but found that it could not fit the
observed excitation-energy dependence. Since the Auger effect is
entirely non-radiative, its presence should also result in a
saturated quantum yield versus excitation intensity, which in our
case starts to appear only above about 1 mJ of excitation energy. We
may conclude that the role of Auger effect is negligible in our
experiments, at least for pulse energies below 1 mJ. This result
allows us to set the following approximate upper limit for the Auger
recombination coefficient: $C_A{\lesssim}10^{-33}$ cm$^6$s$^{-1}$.
This value is two-three orders of magnitude smaller than the typical
values of the Auger coefficient found in indirect semiconductors,
such as germanium or silicon.\cite{zarrabi85,linnros98} The reason
is probably the larger band gap of STO, as the Auger recombination
coefficient is known to decrease exponentially with increasing gap
width.\cite{masse07}

\begin{figure}
\includegraphics[angle=0, width=0.48\textwidth]{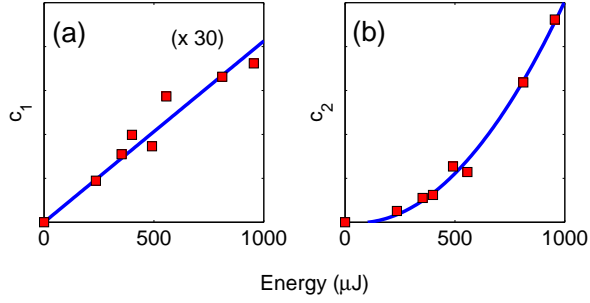}
\caption{(Color online) Amplitude coefficients $c_1$ and $c_2$,
versus excitation pulse energy (proportional to the energy density
$U$) of the unimolecular (a) and bimolecular (b) terms in the
luminescence decay. Square dots are the values extracted from data.
Solid lines correspond respectively to a linear (a) and quadratic
(b) best-fit. Vertical scales of (a) and (b) are the same, but $c_1$
data are multiplied by a factor of 30.\label{fig4}}
\end{figure}

In our BD+UD model, the best-fit values of $\gamma\alpha_2$ were
found to range from 5 to 10 $\mu$s$^{-1}$cm$^3$/J, depending on the
sample (corresponding to a BD time $\tau_2=300-600$ ps for a UV
pulse energy of 1 mJ). Since the ratio $\alpha_2/\alpha_1$ is
unknown, we cannot determine the value of $\gamma$ from this result.
However, a plausible order-of-magnitude estimate is obtained by
assuming a balanced branching ratio $\alpha_2/\alpha_1\approx1$,
which yields $\gamma\approx10^{-11}$ cm$^3$s$^{-1}$ (by a similar
reasoning, we can set the following, well defined, lower limit:
$\gamma>0.5\times10^{-11}$ cm$^3$s$^{-1}$). This value is two-three
orders of magnitude larger than the bimolecular coefficient for
indirect-gap semiconductors such as silicon, usually ascribed to
phonon-assisted radiative and defect-assisted Auger recombination
processes,\cite{linnros98} but it is about one-two orders of
magnitude smaller than the value found in direct band-gap
semiconductors.\cite{zarrabi85,jursenas96} The best-fit values of
$\tau_1$ ranged from 11 to 24 ns. The coefficients $c_1$ and $c_2$
giving the initial amplitude of the UD and BD terms in the
luminescence were adjusted separately for each excitation energy.
The resulting best-fit values are shown in Fig.~\ref{fig4}, where
the energy dependence predicted by Eqs.~(\ref{coeff}) is seen to be
well verified in the non-saturated regime. Besides further
confirming the validity of our model, this shows that both decay
channels are non-saturated in this energy range. For higher
energies, we observe the onset of saturation in both channels (data
not shown). The measured yield ratio of the BD to UD channels is of
$0.6-0.7$, with 20\% sample-to-sample variations but independent of
excitation intensity, within our experimental uncertainties.

Finally, we tried to resolve possible spectral differences between
our UD and BD processes by inserting additional filters in the
detection line and checking for variations in the relative yields or
decay times of UD and BD components. In particular we used a
bandpass filter for the range 390-480 nm, which therefore selects
only the blue section of the spectrum, and a long-pass filter with a
cutoff wavelength at 495 nm, selecting only the green tail of the
spectrum. In both cases, no significant changes were found,
indicating that the emission spectra of UD and BD processes are
largely overlapping.

\section{Discussion}
Let us now discuss the possible underlying nature of the UD and BD
contributions seen in our data. We start from the consideration that
the small photoluminescence absolute yield implies that most
electron-hole recombination processes occurring in STO are
non-radiative, i.e., the radiative transitions that give rise to the
luminescence represent only a small fraction of the total. This in
turn implies that for a given homogeneous population of excited
electrons and holes in STO, the photo-luminescence signal kinetics
is mainly controlled by non-radiative decay channels, regardless of
the fact that it is actually detected through the radiative ones. On
the other hand, the spectrum of the photoluminescence characterizes
only the radiative processes. Therefore, one must be careful when
trying to fit kinetic and spectral pieces of evidence in a single
scenario, as they are likely associated with distinct physical
processes.

As we mentioned in the introduction, charge carriers in STO are
believed to have a polaronic nature, owing to the strong
polarizability of the STO ionic lattice. Electron carriers in STO
are very mobile, especially at low temperatures, as expected from
large polarons,\cite{yasunaga68,keroack84,itoh06,thiel06} although
there is some evidence that localized small polarons also exist in
the system.\cite{eagles96,hasegawa01,yu05} The intrinsic mobility of
hole polarons is unknown, as at low densities STO holes are usually
trapped in mid-gap impurities. Therefore the observed BD kinetics is
to be probably ascribed to the direct non-radiative recombination of
conduction-band electrons and valence-band holes, with at least one
of the two species remaining mobile and delocalized. This
recombination process may be either phonon- or defect-assisted. We
emphasize that the intrinsic BD behavior of the STO luminescence
reported here is quite different from that reported in the
past.\cite{leonelli86,hasegawa00} The latter occurs on a much longer
($\mu$s-to-ms) time-scale and at much smaller charge-carrier
densities and is usually ascribed to carrier trapping-untrapping
processes, dominated by extrinsic effects.\cite{footnote2}

Focusing now on the radiative processes involved in the BD signal,
we must ask ourselves why the luminescence spectrum peak is
red-shifted by about 0.35 eV with respect to the (indirect) band-gap
(Stokes' shift) and why the luminescence band extends down to more
than 1 eV below the excitation energy. Both energies are far too
large to be associated with the emission of a single phonon in a
phonon-assisted inter-band recombination process (the highest-energy
phonon in STO has an energy of about 0.1 eV). Midgap impurities also
cannot be invoked to explain this energy shift, as the number of
estimated $e-h$ pairs well above $10^{20}$ cm$^{-3}$ exceeds by
orders of magnitude the impurity density, so that the impurity
emission would be saturated, in contrast to our results.
Furthermore, the expected sample-to-sample yield fluctuation would
be much larger. If small polarons are indeed formed in STO, the
involved lattice relaxation could instead account for the 0.35 eV
Stokes shift of the blue luminescence peak.\cite{yu05} However,
intrinsic lattice defects such as oxygen vacancies or dislocations
may also play an important role in the radiative processes, either
by introducing mid-gap electronic states or by locally enhancing the
polarizability of the lattice, thus allowing the polaron-like
self-localization of charge carriers.

To account for the UD contribution, we must assume that a fraction
of photo-generated carriers is rapidly converted into bound pairs of
electrons and holes, which then decay more slowly with the
characteristic lifetime $\tau_1$ of the bound pair. Due to the
indirect nature of the gap, $e-h$ pair binding is difficult unless
localization takes place. Again, it is not understood whether this
localization is a purely intrinsic phenomenon or it is somehow
defect-assisted, although the latter hypothesis seems more
plausible, as we find significant sample-to-sample fluctuations in
the UD component yield and lifetime $\tau_1$. If lattice deformation
is instrumental in the localization (both in the perfect crystal or
close to a defect), the localized bound pair may be identified with
the so-called ``self-trapped vibronic exciton''
(STVE).\cite{eglitis03,nasu03,grigorjeva04,vikhnin02} STVE has been
often invoked in the literature to explain the strong luminescent
emission peaking in the green (at 2.4 eV) seen from STO at low
temperature (below 40
K),\cite{grabner69,aguilar82,leonelli86,hasegawa00,kan05}, although
this interpretation has been recently
criticized.\cite{mochizuki05,mochizuki06} In our case, since the
spectrum of the UD component seems roughly similar to the total
spectrum, the STVE should be associated with an emission peaking in
the blue, as was also recently proposed for the case of
electron-induced luminescence in STO.\cite{grigorjeva04}

\section{Conclusions}
Summarizing, the time-resolved luminescence decay observed from STO
under intense pulsed UV excitation is well explained by a model that
assumes the presence of two separate decay channels, one associated
with a direct recombination of unbound mobile electron-hole charge
carriers and the other with the recombination of bound electron-hole
pairs. The exact nature of the microscopic recombination processes
taking place in this system remains uncertain, although the
polaronic character of the charge carriers and the presence of
intrinsic crystal defects are both likely ingredients of a future
understanding. Our experiment also demonstrates the possibility of
probing this photoinduced high charge-density regime without
damaging the sample, thus opening the way to several other possible
investigations of this interesting regime, which may eventually help
us clarifying the exact nature of the charge carriers occurring in
these ever-surprising materials.

\acknowledgments We thank F. Ciccullo for assistance in some
experimental runs.


\begin{thebibliography}{30}
\bibitem{footnote2}
It cannot be excluded, but seems unlikely, that the two phenomena
would merge after taking into account the temperature behavior. Our
preliminary measurements of this behavior do not show a significant
increase of the BD times down to 90 K. Moreover, the blue-peaked
luminescence we are studying in this work corresponds to the ``A
band'' reported in Ref.~\protect\onlinecite{mochizuki05}, where it
is shown to decay faster than 50 ns down to 13 K.
\end{thebibliography}

\end{document}